\documentstyle[aps,prl,multicol,amssymb]{revtex}
\title{Reversibility of local transformations of multiparticle entanglement}

\author{ N.~Linden$^1$,
S. Popescu$^{2,3}$, B.~Schumacher$^4$ and M.~Westmoreland$^{5}$}

\address{
$^1$Department of Mathematics, University of Bristol, University
Walk, Bristol BS8 1TW, UK\\
$^2$H.H. Wills Physics Laboratory,
University of Bristol, Tyndall Avenue, Bristol BS8 1TL, UK\\
$^3$BRIMS, Hewlett-Packard Laboratories, Stoke Gifford, Bristol
BS12 6QZ, UK \\ $^4$Department of Physics, Kenyon College,
Gambier, OH  43022, USA\\ $^5$Department of Mathematical Sciences,
Denison University, Granville, OH  43023, USA}

\date{07 December 1999}
\begin{document}
%\draft
\maketitle
\begin{abstract}
We consider the transformation of multisystem entangled states by
local quantum operations and classical communication.  We show
that, for any reversible transformation, the relative entropy of
entanglement for any two parties must remain constant.  This
shows, for example, that it is not possible to convert $2N$
three-party GHZ states into $3N$ singlets, even in an asymptotic
sense.  Thus there is true three-party non-locality (i.e. not all
three party entanglement is equivalent to two-party entanglement).
Our results also allow us to make {\em quantitative} statements
about concentrating multi-particle entanglement.  Finally, we show
that there is true $n$-party entanglement for any $n$.

\end{abstract}


%\begin{multicols}{2}
\newcommand{\tr}{\mbox{Tr} }
\newcommand{\ket}[1]{\left | #1 \right \rangle}
\newcommand{\bra}[1]{\left \langle #1 \right |}
\newcommand{\amp}[2]{\left \langle #1 \left | #2 \right. \right \rangle}
\newcommand{\proj}[1]{\ket{#1} \! \bra{#1}}
\newcommand{\ave}[1]{\left \langle #1 \right \rangle}
\newcommand{\superop}{{\cal E}}
\newcommand{\unity}{\mbox{\bf I}}
\newcommand{\hilbert}{{\cal H}}
\newcommand{\relent}[2]{S \left ( #1 || #2 \right )}
\newcommand{\banner}[1]{\bigskip \noindent {\bf #1} \medskip}

\begin{multicols}{2}
One of the key open issues in quantum information theory is to
understand what the fundamentally different types of quantum
entanglement are.  It has been known for some time
\cite{purification} that any pure entangled state of two parties,
Alice and Bob, may be reversibly distilled to singlets in the
sense that in the limit of large $N$, $N$ copies of the state
$\psi^{AB}$ may be transformed reversibly into $N E(\psi^{AB})$
singlets, where $ E(\psi^{AB})$ is the entropy of the reduced
density matrix of either Alice or Bob. Until recently it was not
known whether, in fact, singlets are the only type of
entanglement.  This issue was resolved in \cite{Bennett99} where
was shown that certain multi-party states cannot be transformed
into singlets reversibly. In particular the authors of
\cite{Bennett99} consider the four-party GHZ state:
\begin{eqnarray}
   \psi_{ GHZ_4} ={1\over \sqrt 2}\left( | 0000 \rangle + | 1111 \rangle
   \right).\label{GHZ4}
  \end{eqnarray}
Let us imagine that there were a reversible process to convert $N$
copies of $\psi_{GHZ_4}$ into singlets (in the limit of large
$N$). After the forward version of the process there would be
$s^{AB}$ singlets between Alice and Bob, $s^{BC}$  singlets
between Bob and Claire etc.  It is straightforward to calculate
the four one-party entropies (e.g. the entropy of Alice versus the
Bob-Claire-Daniel system) and three independent two-party
entropies (e.g. the entropy of the Alice-Bob system versus the
Claire-Daniel system) for both the initial and final pure states.

Since entropy can only decrease during any local process, it must
be constant during a reversible process.  Thus a {\em necessary}
condition for the existence of a reversible protocol for
converting $N$ copies of $\psi_{GHZ_4}$ into singlets is that the
entropies of the initial and final states must be the same. It is
not difficult to show that no combinations of singlets held
between the four parties has the same ratios of entropies as
$\psi_{GHZ_4}$. Thus the four-party GHZ state cannot be converted
reversibly into singlets.  Hence not all four-party entanglement
is of the singlet type.

\cite{Bennett99} leaves open many questions. For example, while
the results in \cite{Bennett99} show that not all four-party
entanglement can be attributed to pair-wise entanglement, the
techniques employed leave open whether or not any new types of
non-locality arise in three-party states. Consider the three
one-party entropies, $E^A,\ E^B,\ E^C$ of an arbitrary three party
pure state.  It can easily be checked that any values of these
three entropies which are allowed by sub-additivity can be matched
by suitable choices of the numbers of singlets held between the
three parties.

A particularly important case is that of the three party GHZ state
\begin{eqnarray}
   \psi_{GHZ_3} ={1\over \sqrt 2}\left( | 000 \rangle + | 111 \rangle
   \right).\label{GHZ3}
  \end{eqnarray}
A simple calculation shows that  the one-party entropies of
$\psi_{GHZ_3}$ can be matched by having one singlet between each
pair $AB$, $BC$, $AC$ for every two $\psi_{GHZ_3}$'s held between
the parties.  There is further encouragement for the suggestion
that, in fact, the three party GHZ state is equivalent to
singlets, since it is  possible to create singlets between
any pair from $\psi_{GHZ_3}$. To see this rewrite (\ref{GHZ3})
as
 \begin{eqnarray}
   \psi_{GHZ_3}&=&{1\over \sqrt 2}\bigg[
   {\left( | 0 \rangle + | 1 \rangle \right)\over \sqrt 2}
   {\left( | 00 \rangle + | 11 \rangle \right)\over \sqrt 2}
\\ \nonumber
& &\qquad +
 {\left( | 0 \rangle - | 1 \rangle \right)\over \sqrt 2}
 {\left( | 00 \rangle - | 11 \rangle \right)\over \sqrt 2}
 \bigg].
 \end{eqnarray}
Now let Alice measure her particle in the $ x$-basis,

\begin{eqnarray}
|+\rangle_x =  {\left( | 0 \rangle + | 1 \rangle \right)\over \sqrt 2};\qquad
|-\rangle_x =  {\left( | 0 \rangle - | 1 \rangle \right)\over \sqrt 2}.
\end{eqnarray}
If she
finds her particle in the $+x$ direction she tells Bob and Claire
to do nothing, if she finds her particle in the $-x$ direction she
tells Bob to do the unitary transformation on his particle (a
rotation about the $z$-axis)
  \begin{eqnarray}
| 0 \rangle \mapsto | 0 \rangle;\qquad | 1 \rangle \mapsto -|1
\rangle
  \end{eqnarray}
  At the end of these operations Bob and Claire share the state,
  \begin{eqnarray}
   {1\over \sqrt 2}\left( | 00 \rangle + | 11 \rangle
   \right).
  \end{eqnarray}
Thus given one three party GHZ state, one singlet  between Bob and
Claire can be produced.  However the only known protocol for
converting singlets to GHZ states consumes two singlets; for
example given singlets between Alice and Bob and between Bob and
Claire, Bob could create a GHZ state locally and then use the
singlets to  teleport to Alice and Claire. Thus while individual
copies of GHZ states can be converted to singlets and vice versa,
the only known protocol is not reversible. Indeed it has been
shown that no reversible protocol exists for finite numbers of
copies \cite{Bennett99}. A natural question is whether in the
asymptotic limit any such reversible protocol can exist. Our
results below settle the question.

In this letter we first investigate the three-party case.  We will
consider general collective actions on any number, $N$, of copies
of the state (including the asymptotic limit
$N\rightarrow\infty$). We derive new conditions that any procedure
involving local quantum operations and classical communication
must satisfy, namely that the increase in the average relative
entropy of entanglement of any pair of the three parties cannot be
greater than the decrease in the entanglement of the third with
the pair. A corollary of this result is  that the average relative
entropy of entanglement of any pair of the parties must be
constant in a reversible process. This allows us easily to
conclude that not all three party entanglement is of singlet type
since we can show that there is no reversible process converting
the three party GHZ state into singlets, even in the asymptotic
limit.  More generally we will show that there is true $n$-party
entanglement for any $n$, in the sense that, for any $n$ there are
$n$-party states which cannot be transformed reversibly into
states in which only $k<n$ parties are entangled.

 It will simplify matters if we consider the entire Hilbert space
at each site including ancilla's.  Thus initially Alice, Bob and
Claire share a pure state.  They are allowed to perform local
operations and classical communication, as usual. The state of the
system branches according to the outcome of any measurements
performed, but at each stage the state in a given branch is still
pure.  The set of local operations can be broken down into two
classes:
\begin{itemize}
\item  Alice Bob and Claire can do local unitary operations on their systems,
    and since they
    can communicate, everybody can know what transformations are done
    by everybody else. Local unitary transformations cannot
    increase the entanglement.
\item  They can perform more general operations, including measurements.
    Communication means that everybody can know the results of every
    measurement performed by anybody.

\end{itemize}
We note that entanglement between Bob and Claire can be increased.
For example, a measurement performed by
    Alice can  increase the entanglement of Bob and
    Claire.
    The example of the three-party GHZ state above shows this
    since by taking the trace over Alice's Hilbert space one can see that
Bob and Claire's relative state is unentangled in the case of the
three-party GHZ state. However after the protocol Bob and Claire's
entanglement is one e-bit.  It is relevant, however (see below)
 that after
  the protocol Alice is unentangled with Bob and Claire.

  Our aim below is to calculate how much Bob and Claire's
entanglement can increase under the most general local operations
that the three parties can perform.

 Specifically we will show that
any increase in the Bob-Claire bipartite entanglement must be paid
for by a decrease in the entanglement of Alice with the other
two---that is (since we always have a pure state) by a decrease in
Alice's entropy.

The proof is as follows. Consider an entanglement manipulation
protocol which states with any number of copies of a three-party
entangled state.  Consider a particular stage in the protocol and
a particular branch in which the density matrix of the system is
$\rho^{ABC}$ (note that the state is pure, and lives in the
Hilbert space of all the original copies including the ancillas);
the state of Bob and Claire's joint system is
$\rho^{BC}=\tr_A\left(\rho^{ABC}\right)$. Let Alice perform a
measurement of an operator with spectral projectors $P_k$.  Thus
if the outcome $k$ is obtained, the state of the system is
  \begin{eqnarray}
{ P_k \otimes \unity \otimes\unity \rho^{ABC} P_k \otimes
\unity\otimes\unity \over \tr \left( P_k \otimes \unity
\otimes\unity \rho^{ABC}\right)},
\end{eqnarray}
where
$
p_k = \tr \left( P_k \otimes \unity \otimes\unity
\rho^{ABC}\right)
$
is the probability that the outcome $k$ is obtained. We note that
 \begin{eqnarray}
 \rho^{BC}= \sum_k p_k  \rho^{BC}_k,
\end{eqnarray}
where $\rho^{BC}_k$ is the state of Bob and Claire's joint system
after the measurement; in other words after Alice's measurement,
but before she has communicated the outcome to Bob and Claire,
their average state cannot have changed from what it was before
the measurement.

The relative entropy of $\rho^{BC}$ with respect to any bipartite
state $\sigma^{BC}$ is defined as
 \begin{eqnarray}
\relent{\rho^{BC}}{\sigma^{BC}} &:&=\\ \nonumber
& &
\tr\left(\rho^{BC}\ln\rho^{BC}\right)
 - \tr\left(\rho^{BC}\ln \sigma^{BC}\right)
\end{eqnarray}

Simple algebra shows that the relative entropy satisfies
``Donald's identity'' \cite{Donald},
\begin{eqnarray}
  & &  - \relent{\rho^{BC}}{\sigma^{BC}}\\ \nonumber
    & &\qquad    =  \sum_{k} p_{k} \left (
        \relent{\rho^{BC}_{k}}{\rho^{BC}} -
        \relent{\rho^{BC}_{k}}{\sigma^{BC}}  \right ) .
\end{eqnarray}
The relative entropy of entanglement \cite{PlenioVedral}
for $\rho^{BC}$ is defined as
\begin{equation}
    E_{r}\left(\rho^{BC}\right)
        = \min_{\sigma^{BC} \mbox{\tiny sep}}
            \relent{\rho^{BC}}{\sigma^{BC}}
\end{equation}
where we've minimized over separable $\sigma^{BC}$. From now on,
let us choose $\sigma^{BC}$ to be the separable density operator
that does this minimization for Bob and Claire's state
$\rho^{BC}$. Then we obtain
\begin{equation}
    \sum_{k} p_{k} \relent{\rho^{BC}_{k}}{\sigma^{BC}}
        - E_{r}\left(\rho^{BC}\right)  =
        \sum_{k} p_{k} \relent{\rho^{BC}_{k}}{\rho^{BC}} ,
\end{equation}
which means that
\begin{eqnarray}
 & &   \sum_{k} p_{k} E_{r}\left(\rho^{BC}_{k}\right)
    - E_{r}\left(\rho^{BC}\right)  \\ \nonumber
    & &\qquad\qquad  \leq
        \sum_{k} p_{k} \relent{\rho^{BC}_{k}}{\rho^{BC}} \\ \nonumber
   & &\qquad\qquad  =  S\left(\rho^{BC}\right) -
        \sum_{k} p_{k} S\left(\rho^{BC}_{k}\right) \\ \nonumber
    &  &\qquad\qquad  = S\left(\rho^{A}\right) -
        \sum_{k} p_{k} S\left(\rho^{A}_{k}\right);
\end{eqnarray}
the last step is valid since  $\rho^{ABC}$ and $\rho^{ABC}_k$
are pure.

Now we consider a further step in which Alice communicates to Bob
and Claire the results of her measurement and then Bob and Claire
perform unitary rotations on their states with operators which
depend on the outcome; we also allow Alice to perform rotations on
her state which depend on the outcome of the measurement. We will
denote by $\tilde\rho^{BC}_{k}$ and $\tilde\rho^{A}_{k}$ the
states after the transformations i.e.
\begin{eqnarray}
 & &  \rho^{BC}_{k}\mapsto
    \tilde \rho^{BC}_{k}= U^B_k\otimes U^C_k \rho^{BC}_k
   \otimes \left(U^C_k\right)^\dagger;\\ \nonumber
& &\rho^{A}_{k}\mapsto \tilde \rho^{A}_{k}=U^A_k
\rho^{A}_k\left(U^A_k\right)^\dagger.
\end{eqnarray}
These transformations do not change $S\left(\rho^{A}_{k}\right)$.
 $E_{r}\left(\rho^{BC}_{k}\right)$ is also left unchanged by these
transformations
 since the set of separable states is invariant under local
 unitary transformations  (i.e if $\sigma^{BC}$ is separable, then so is
$U\otimes V\sigma^{BC}U^\dagger\otimes V^\dagger$) so that we find
\begin{eqnarray}
  & &  \sum_{k} p_{k} E_{r}\left(\tilde\rho^{BC}_{k}\right)
    - E_{r}\left(\rho^{BC}\right)  \label{inequalityone} \\ \nonumber
    & &\qquad\qquad  \leq
         S\left(\rho^{A}\right) -
        \sum_{k} p_{k}
        S\left(\tilde\rho^{A}_{k}\right);
\end{eqnarray}
Thus, the average increase in $E_{r}$ for the Bob-Claire system is
no greater than the average decrease in the entropy of Alice's
system (and thus Alice's entanglement with the joint Bob-Claire
system).

Inequality (\ref{inequalityone}) is also true in a step of an
extended protocol in which Bob performs a measurement communicates
to Alice and Claire, and then all three parties perform unitary
transformations dependent on the outcome of the measurement. For
again we consider a particular stage in the protocol and a
particular branch in which the density matrix of the system is
$\rho^{ABC}$ (recall that the state is pure). Let Bob perform a
measurement of an operator with spectral projectors $P_k$.  Thus,
as before,
 if the outcome $k$ is obtained, we denote the state of the system
$
  \rho^{ABC}_k$ and $
 p_k $
is the probability that the outcome $k$ is obtained. Alice's
reduced state after the measurement is
 \begin{eqnarray}
\rho^{A}_k  = \tr_{BC} \left( \rho^{ABC}_k\right).
\end{eqnarray}
Before Bob communicates to her, her average state is
 \begin{eqnarray}
\rho^{A} = \sum p_k \rho^{A}_k  .
\end{eqnarray}
Thus the convexity of entropy shows that
\begin{eqnarray}
 S\left(\rho^{A}\right) \geq
        \sum_{k} p_{k}
        S\left(\rho^{A}_{k}\right).
\end{eqnarray}
Now Bob communicates the outcome of his measurement and Alice
performs a unitary transformation which depends on this outcome:
\begin{eqnarray}
\rho^{A}_{k}\mapsto \tilde \rho^{A}_{k}=U^A_k
\rho^{A}_k\left(U^A_k\right)^\dagger.
\end{eqnarray}
These transformations do not change $S\left(\rho^{A}_{k}\right)$.
Thus, during this step of the protocol,
\begin{eqnarray}
 S\left(\rho^{A}\right) \geq
        \sum_{k} p_{k}
        S\left(\tilde\rho^{A}_{k}\right).
        \label{entropyinequality}
\end{eqnarray}

It is a key property of the relative entropy of entanglement that
it does not increase under local operations and classical
communication (see for example \cite{PlenioVedral}) so that
\begin{eqnarray}
E_r\left(\rho^{BC}\right) \geq
        \sum_{k} p_{k}
       E_r\left(\tilde\rho^{BC}_{k}\right).
        \label{relativeentropyinequality}
\end{eqnarray}

Thus (\ref{entropyinequality}) and
(\ref{relativeentropyinequality}) imply that (\ref{inequalityone})
is true for any step in the protocol.

 If we consider an extended protocol in
which Alice, Bob and Claire perform many rounds of local
measurement, classical communication and unitary transformations,
we may apply the above inequality to each round for each branch.
We can then deduce that for any local protocol, the average
increase in $E_{r}$ for the Bob-Claire system is no greater than
the average decrease in the Alice's entanglement with the joint
Bob-Claire system. i.e. we may write
\begin{eqnarray}
  & & \left \langle E_{r}\left({BC}\right) \right \rangle_{\rm final}
    - E_{r}\left({BC}\right)_{\rm initial}  \label{overallinequality}
    \nonumber \\
    & & \qquad \qquad \leq
         S\left({A}\right)_{\rm initial} -
        \left \langle S\left({A}\right) \right \rangle_{\rm final}.
\end{eqnarray}

The general question we are interested in is reversible procedures
for converting a given state to some specified states.  For a
reversible process we know that average entropy cannot change
since entropy can only stay constant or decrease under local
operations. Thus in a reversible process the right-hand-side of
(\ref{overallinequality}) is zero and so in such a process  the
average relative entropy of entanglement of  the Bob-Claire system
must be constant.

This form of the result makes it very easy to show that  it is not
possible to convert GHZ states reversibly into singlets between
Alice and Bob, Bob and Claire, and Alice and Claire.  This is
because the relative entropy of entanglement of GHZ states between
Bob and Claire is zero, but clearly singlets between Bob and
Claire have non-zero relative entropy of entanglement.

The results above lead us to our second main point namely to
make quantitative statements about multi-party entanglement.
Having established that GHZs and singlets are not interconvertible
it is natural to ask, following \cite{Bennett99} whether three
party pure states can be transformed reversibly into singlets {\em
and} three-party GHZ's. That is, perhaps singlets and three-party
GHZ's constitute the irrreducible types of entanglement into which
any three party entanglement can be transformed reversibly. We do
not know whether this is possible in general. However for those
states $\phi^{ABC}$ for which it is possible, our results easily
show how many singlets and GHZ's can be extracted. This is since
the one-party entropies and the relative entropy of entanglement
of the reduced two-party density matrices are conserved as we have
shown.  Thus if $g$ is the number of GHZ's that can be extracted
per individual copy of $\phi^{ABC}$, and if $s_{AB}$, $s_{BC}$,
and $s_{AC}$ are the number of singlets, then
\begin{eqnarray}
  S_A(\phi^{ABC}) &=& g + s_{AB}+ s_{AC}  \label{numbersextractable}
        \\ \nonumber
 S_B(\phi^{ABC}) &=& g + s_{AB}+ s_{BC}  \\ \nonumber
 S_C(\phi^{ABC}) &=& g + s_{AC}+ s_{BC}  \\ \nonumber
 E_r(\rho^{AB}) & =  &s_{AB}\\ \nonumber
 E_r(\rho^{BC}) & =  &s_{BC}\\ \nonumber
 E_r(\rho^{AC}) & =  &s_{AC,}
 \end{eqnarray}
where $\rho^{AB}$ etc. are the reduced density matrices of
$\phi^{ABC}$. i.e. the number of singlets between each pair $AB$,
$BC$, $AC$ that can be extracted per state asymptotically is equal
to the relative entropy of entanglement of the reduced density
matrices. We note that (\ref{numbersextractable}) shows that for
states which are convertible into GHZ's and singlets, there are
relations between the one-party entropies and relative entropies.

One interesting case is the state
\begin{eqnarray}
   \phi_1 =\alpha | 000 \rangle +\beta | 111 \rangle.
  \end{eqnarray}
The conservation laws above suggest that the number of GHZ's that
can be extracted is equal to
\begin{eqnarray}
H(\alpha^2)=- \alpha^2 \log \alpha^2 -\beta^2 \log \beta^2.
\end{eqnarray}
Indeed a simple extension of the standard purification protocol
\cite{purification} shows that this is indeed possible.

A second interesting case is
\begin{eqnarray}
   \phi_2 =\alpha | 0 \rangle \Psi_{+} +\beta | 1 \rangle\Psi_{-},
  \end{eqnarray}
where $ \Psi_{\pm}= {1\over\sqrt 2}\left(| 00 \rangle \pm  11
\rangle \right)$.  It is very tempting to think that this state
can indeed be transformed asymptotically into GHZ's and singlets,
and the arguments above lead to a conjecture for the numbers of
these states which can be extracted, namely $H(\alpha^2)$
GHZ's and $1-H(\alpha^2)$ singlets between Bob and Claire,
per copy of $\phi_2$.
 At present  there is no protocol known to perform this
 transformation.

Our argument can be extended to situations in which multisystem
entanglement is shared among more than three separated parties.
Suppose $n$ parties share $n$ quantum systems in a joint entangled
state.  As long as $n \geq 3$, we can partition the $n$ parties
into three non-empty groups, which will play the roles of Alice,
Bob and Claire. Local operations by any of the $n$ parties will
necessarily be local operations with respect to the
Alice/Bob/Claire partition. Any increase in the relative entropy
of entanglement between the Bob and Claire groups due to
operations by the Alice group will necessarily involve an
irreversible reduction in the entropy of the Alice group, and thus
a reduction in the entanglement of the Alice group with the others.

Any entangled pure state of $k$ parties must show bipartite
entanglement between some subset of the $k$ parties and the
complementary subset.  This fact allows us to draw conclusions
about the reversible transformations of many-particle
entanglement. For instance, a GHZ state shared among $n$ parties
has the property that for $k < n$, no $k$ parties are entangled
among themselves.  Thus, $n$-party GHZ's cannot be reversibly
transformed into any combination of $k$-party entangled pure
states, for all $k < n$.

Finally we point out that in our derivation of
(\ref{overallinequality}), we had in mind the definition of
relative entropy of entanglement given by
\begin{equation}
    E_{r}\left(\rho^{BC}\right)
        = \min_{\sigma^{BC} \mbox{\tiny sep}}
            \relent{\rho^{BC}}{\sigma^{BC}}.
\end{equation}
That is, the set of states $\Sigma$ over which we minimize is the
set of separable states.  However, the only property of that set
necessary for the proof was the {\em invariance} of that set under
local transformations, that is unitary transformations and
measurements (in particular \cite{PlenioVedral} the invariance
under measurements enters in the derivation of
(\ref{relativeentropyinequality})). For two parties, a natural
such set is the set of separable states. For larger numbers of
parties, more general choices are possible (see for example
\cite{PlenioVedral}), and can provide new measures of multisystem
entanglement.

For example for four parties, we can take the states in $\Sigma$
to be mixtures of pure states of the form
$
  \ket{\psi^{ABC}} \otimes
                \ket{\phi^{D}},
$
or we could take mixtures of these states and similar states with
$ABCD$ permuted. Or we could take $\Sigma$ to be mixtures of pure
states of the form
 $
  \ket{\psi^{AB}} \otimes
                \ket{\phi^{CD}}$ and permutations etc.
In any of these cases the set $\Sigma$ is invariant under local
transformations. We can therefore use it to define
$E_{\Sigma}(\rho^{ABCD})$, the relative entropy ``distance'' of a
state $\rho^{ABCD}$ from the set $\Sigma$. Similar reasoning to
that given earlier allows us to derive inequalities similar to
(\ref{inequalityone}):
\begin{eqnarray}
  & &  \sum_{k} p_{k} E_{\Sigma}\left(\tilde\rho^{BCDE}_{k}\right)
    - E_{\Sigma}\left(\rho^{BCDE}\right)  \nonumber \\
    & &\qquad\qquad  \leq
         S\left(\rho^{A}\right) -
        \sum_{k} p_{k}
        S\left(\tilde\rho^{A}_{k}\right);\label{fourpartyinequality}
\end{eqnarray}

More generally for any $n$ we can consider the set $\Sigma$ to be
states which are mixtures of pure states with any given
partitioning of all the parties. A heirarchy of entanglement
measures emerges, each member of which must be conserved in
reversible transformations.

We are very grateful to Serge Massar for discussions and for
pointing out a mistake in an early version of the manuscript. This
work was made possible by the Isaac Newton Institute programme on
``Complexity, Computation and the Physics of Information'' (1999),
partly supported by the European Science Foundation.  One of us
(BS) also acknowledges the support of a Rosenbaum Fellowship
during this programme.

\end{multicols}

\begin{thebibliography}{10}
\bibitem{purification} C.H. Bennett, H. Bernstein, S. Popescu and
B.W. Schumacher, Phys. Rev. A53 (1996) 3824
\bibitem{Bennett99}C.H.
Bennett, S. Popescu, D. Rohrlich, J.A. Smolin and A.V. Thapliyal,
quant-ph/9908073.
\bibitem{Donald}  M. Ohya and D. Petz, {\em Quantum Entropy and Its Use},
(Springer-Verlag, Berlin, 1993).
\bibitem{PlenioVedral}V. Vedral and  M. Plenio, Phys. Rev. A57
(1998) 1619.
\end{thebibliography}
\end{document}